\documentclass[12pt]{article}
\usepackage{amsfonts}
\usepackage{amssymb}
\usepackage{amscd}

\begin{document}
%%    The information for the title page will be placed between
%%    \begin{document} and \maketitle. The order of most entries
%%    is determined by the class file and can not be changed by
%%    rearranging them. The maketitle command follows after the
%%    abstract.
%%
%%    Most of the following commands will be completed by the publisher.
%%
%%    The copyrightyear is defined in the .clo file as the first argument
%%    of the copyrightinfo command. If the copyrightyear differs from that
%%    value it might be adjusted by the following definition:
%%
%% \renewcommand{\copyrightyear}{2003}% uncomment to change the copyrightyear.
%%
%\DOIsuffix{theDOIsuffix}
%%
%% issueinfo for header and copyright line
%\Volume{XX} \Issue{X} \Month{XX} \Year{2009} %%
%%    First and last pagenumber of the article. If the option
%%    'autolastpage' is set (default) the second argument may be left empty.
%\pagespan{1}{} %%
%%    Dates will be filled in by the publisher. The 'reviseddate' and
%%    'dateposted' (Published online) entry may be left empty.
%\Receiveddate{15 November 2007}
%\Reviseddate{30 November 2007}
%\Accepteddate{2 December 2007}
%\Dateposted{3 December 2007}
%%
%\keywords{Effective Lagrangians, $p$-adic strings, adelic strings,
% zeta function, nonlocal field theory.}
% \subjclass[pacs]{11.10.Lm, 11.25.-w}

%% \pretitle{Editor's Choice}

%% We have a short and a long form for the title. The short form
%% (optional argument) goes into the running head.

\title{\bf Towards effective Lagrangians \\ for adelic strings}

%% Please do not enter footnotes or \inst{}-notes into the optional
%% argument of the author command. The optional argument will go into
%% the header.  If there is only one address the marker \inst{x} may be
%% omitted.

%% Information for the first author.
\author{Branko Dragovich
  \footnote{ E-mail:~\textsf{dragovich@phy.bg.ac.yu}} \\ {} \\
  {\it Institute of Physics} \\
 {\it Pregrevica 118, P.O. Box 57,  11001
Belgrade, Serbia}} %%
%%    Information for the second author
%\author[S. Author]{Second Author\inst{1,2,}\footnote{Second author footnote.}}
%\address[\inst{2}]{Second address}
%%
%%    Information for the third author
%\author[T. Author]{Third Author\inst{2,}\footnote{Third author footnote.}}
%%
%%    \dedicatory{This is a dedicatory.}
\date{~}

\maketitle

\begin{abstract}
$p$-Adic strings are important objects of string theory, as well
as of $p$-adic mathematical physics and nonlocal cosmology. By a
concept of adelic string one can unify and simultaneously study
various aspects of ordinary and $p$-adic strings. By this way, one
can consider adelic strings as a very useful instrument in the
further investigation of modern string theory.  It is remarkable
that for some scalar $p$-adic strings exist effective Lagrangians,
which are based on real instead of $p$-adic numbers and describe
not only four-point scattering amplitudes but also all higher ones
at the tree level. In this work, starting from $p$-adic
Lagrangians, we consider some approaches to construction of
effective  field Lagrangians for $p$-adic sector of adelic
strings. It yields Lagrangians for nonlinear and nonlocal scalar
field theory, where spacetime nonlocality is determined by an
infinite number of derivatives contained in the operator-valued
Riemann zeta function. Owing to the Riemann zeta function in the
dynamics of these scalar field theories, obtained Lagrangians are
also interesting in themselves.
\end{abstract}
%% maketitle must follow the abstract.
\maketitle                   % Produces the title.

%% If there is not enough space inside the running head
%% for all authors including the title you may provide
%% the leftmark in one of the following three forms:

%% \renewcommand{\leftmark}
%% {F. Author: A short title}

%% \renewcommand{\leftmark}
%% {F. Author and S. Author: A short title}

%% \renewcommand{\leftmark}
%% {F. Author et al.: A short title}

%% \tableofcontents  % Produces the table of contents.
\section{Introduction}

It is well known that string theory is the best candidate for a
unification of fundamental forces and elementary particles. People
is mainly interested  in ordinary strings, which description is
based on real (and complex) numbers. In 1987, some $p$-adic
analogs \cite{volovich1} of ordinary open and closed strings were
introduced. Description of these $p$-adic strings employs, more or
less, $p$-adic numbers. The most popular $p$-adic strings have
only world sheet parameterized by $p$-adic numbers, while all
other ingredients of the scattering amplitude use real (or
complex) numbers. The starting point is the usual crossing
symmetric Veneziano amplitude
\begin{equation}
A_\infty (a, b) = g_\infty^2\, \int_{\mathbb{R}}
|x|_{\infty}^{a-1} \, |1-x|_{\infty}^{b-1}\, d_\infty x =
g_\infty^2 \frac{\zeta (1-a)}{\zeta (a)}\, \frac{\zeta
(1-b)}{\zeta (b)}\, \frac{\zeta (1-c)}{\zeta (c)}\,,
\label{1.1}
\end{equation}
where $a = - \alpha (s) = - \frac{s}{2} - 1,\, b = - \alpha (t),\,
c = - \alpha (u)$ with the condition $a + b + c = 1$, i.e. $s + t
+ u = - 8$. In (\ref{1.1}), $\, |\cdot|_\infty$ denotes the
ordinary absolute value, $\mathbb{R}$ is the field of real
numbers, kinematic variables $a, b, c \in \mathbb{C}$, and $\zeta$
is the Riemann zeta function. The corresponding Veneziano
amplitude for scattering of $p$-adic strings was introduced
\cite{freund1} as $p$-adic analog of the integral form of
(\ref{1.1}), i.e.
\begin{equation}
A_p (a, b) = g_p^2\, \int_{\mathbb{Q}_p} |x|_p^{a-1} \,
|1-x|_p^{b-1}\, d_p x \,, \label{1.2}
\end{equation}
where $\mathbb{Q}_p$ is the field of $p$-adic numbers, $|\cdot
|_p$ is $p$-adic absolute value and $d_p x$ is the Haar measure on
$\mathbb{Q}_p$. In (\ref{1.2}), kinematic variables $a, b, c$
maintain their complex values and condition $a + b + c = 1$. Thus
$x$, which is related to the string world sheet,  is real-valued
argument in (\ref{1.1}) and $p$-adic one in (\ref{1.2}).
Performing integration in (\ref{1.2}) one obtains
\begin{equation}
A_p (a, b) = g_p^2\, \frac{1- p^{a-1}}{1- p^{-a}}\, \frac{1-
p^{b-1}}{1- p^{-b}}\,\frac{1- p^{c-1}}{1- p^{-c}}\,, \label{1.3}
\end{equation}
where $p$ is any prime number. Recall the definition of the
Riemann zeta function
\begin{equation}
\zeta (s) = \sum_{n= 1}^{+\infty} \frac{1}{n^{s}} = \prod_p
\frac{1}{ 1 - p^{- s}}\,, \quad s = \sigma + i \tau \,, \quad
\sigma
>1\,, \label{1.4}
\end{equation}
which has analytic continuation to the entire complex $s$ plane,
excluding the point $s=1$, where it has a simple pole with residue
1.  According to (\ref{1.4}) one has
\begin{equation}
\prod_p A_p (a, b) =  \frac{\zeta (a)}{\zeta (1-a)}\, \frac{\zeta
(b)}{\zeta (1-b)}\, \frac{\zeta (c)}{\zeta (1-c)} \, \prod_p
g_p^2\,, \label{1.5}
\end{equation}
what gives a nice simple product formula \cite{freund2}
\begin{equation}
A_\infty (a, b) \, \prod_p A_p (a, b) = g_\infty^2 \, \prod_p
g_p^2\,, \label{1.6}
\end{equation}
which to be finite requires $g_\infty^2 \, \prod_p g_p^2 = const.$
From (\ref{1.6}) follows that the ordinary Veneziano amplitude,
which is rather complex, can be expressed as product of all
inverse $p$-adic counterparts, which are much more simpler.
Moreover, expression (\ref{1.6}) gives rise to consider it as the
amplitude for an adelic string, which is composed of the ordinary
and $p$-adic ones.

Various $p$-adic structures have been observed and investigated
not only in string theory but also in many other models of modern
mathematical physics and related topics (see
\cite{freund3,volovich2} for a review of early developments, and
\cite{dragovich1} for the most recent short overview of $p$-adic
mathematical physics). For instance, in \cite{dragovich2} was
shown that degeneration of the genetic code has a $p$-adic
structure. For a basic information on $p$-adic numbers one can
refer to \cite{freund3} and \cite{volovich2}.

It is worth pointing out that for an open scalar $p$-adic string
exists an effective Lagrangian, which is based on real instead of
$p$-adic numbers and describes all scattering amplitudes at the
tree level \cite{freund4,frampton1}.

The adelic approach connects $p$-adic phenomena with the related
ordinary ones. In addition to product formula for string
amplitudes (\ref{1.6}), adelic modelling has been successfully
applied to quantum mechanics \cite{dragovich-a}, Feynman path
integral \cite{dragovich-b}, quantum cosmology \cite{dragovich-c},
summation of divergent series \cite{dragovich-d}, and dynamical
systems \cite{dragovich-e}.

The present paper can be regarded as a result of some attempts to
construct an effective Lagrangian for entire  $p$-adic sector of
an adelic scalar string. We use two approaches, which we call
additive and multiplicative. In the additive approach, we start
with the exact Lagrangian for the effective field of $p$-adic
tachyon string, then extend prime number $p$ to arbitrary natural
number $n$, and perform various summations of such Lagrangians
over $n$. It leads us to the possibility to apply the summation
form (\ref{1.4}) of the zeta function. The multiplicative approach
 enables to use the product over primes form (\ref{1.4}) to
introduce the zeta function.

\section{Lagrangian for a $p$-adic open string}

The exact tree-level Lagrangian of the effective scalar field
$\varphi$, which describes the open $p$-adic string tachyon, is
\cite{freund4,frampton1}
\begin{equation} {\cal L}_p = \frac{m_p^D}{g_p^2}\, \frac{p^2}{p-1} \Big[
-\frac{1}{2}\, \varphi \, p^{-\frac{\Box}{2 m_p^2}} \, \varphi  +
\frac{1}{p+1}\, \varphi^{p+1} \Big]\,,  \label{2.1} \end{equation}
where $p$
 is a prime, $\Box = - \partial_t^2  + \nabla^2$ is the
$D$-dimensional d'Alembertian and we adopt the metric with
signature $(- \, + \, ...\, +)$. At the first glance it may look
strange that there is nothing $p$-adic in Lagrangian (\ref{2.1}).
However, (\ref{2.1}) can be rewritten in the form
\begin{equation} {\cal L}_p = \frac{m_p^D}{g_p^2}\, \frac{1}{|p|_p (1-|p|_p)} \Big[
-\frac{1}{2}\, \varphi \, |p|_p^{\frac{\Box}{2 m_p^2}} \, \varphi
+ \frac{|p|_p}{|p|_p+1}\, \varphi^{\frac{|p|_p + 1}{|p|_p}}
\Big]\,, \label{2.2}
\end{equation}
where prime $p$ is treated as a $p$-adic number. Using $p$-adic
norm of $p$ in (\ref{2.2}) gives real prime in (\ref{2.1}). Note
that similarly the Riemann zeta function can be introduced as
\begin{equation}
\zeta (s) =  \prod_p \frac{1}{ 1 - |p|_p^{ s}}\,, \quad s = \sigma
+ i \tau \,, \quad \sigma
>1\,, \label{2.3}
\end{equation}
where its origin in $p$-adics is evident. In the  sequel we shall
use the usual real form (\ref{2.1}) and (\ref{1.4}).

An infinite number of spacetime derivatives follows from the
expansion
$$
p^{-\frac{\Box}{2 m_p^2}} = \exp{\Big( - \frac{1}{2 m_p^2}
\log{p}\, \Box \Big)} = \sum_{k = 0}^{+\infty} \, \Big(-\frac{\log
p}{2 m_p^2} \Big)^k \, \frac{1}{k !}\, \Box^k \,.
$$
The equation of motion for (\ref{2.1}) is

\begin{equation} p^{-\frac{\Box}{2 m_p^2}}\, \varphi = \varphi^p \,,
\label{2.4} \end{equation} and its properties have been studied by
many authors (see, e.g. \cite{zwiebach1,barnaby1,vladimirov1} and
references therein).

Based on (\ref{2.1}), many aspects of  $p$-adic string dynamics
were considered and compared with the dynamics of ordinary strings
(see, e.g. \cite{sen,zwiebach1,arefeva}     and references
therein) and cosmology \cite{barnaby2,dragov1}.

\section{Zeta-nonlocal Lagrangians}

Starting from (\ref{2.1}), we explore two approaches to the
extension of $\mathcal{L}_p$ that lead to introduction of the
Riemann zeta function. We will first review additive approach and
then introduce a multiplicative one.

\subsection{Additive approach}

It is worth noting that prime number $p$ in (\ref{2.1}) and
(\ref{2.2}) can be replaced by any natural number $n \geq 2$ and
such expressions  also make sense.

Now we want to introduce a Lagrangian which incorporates all the
above  Lagrangians (\ref{2.1}), with $p$ replaced by $n \in
\mathbb{N}$. To this end, we take the sum of all Lagrangians
${\cal L}_n$  in the form

\begin{equation} L =   \sum_{n = 1}^{+\infty} C_n\, {\cal L}_n   =
 \sum_{n= 1}^{+\infty} C_n \frac{ m_n^D}{g_n^2}\frac{n^2}{n -1}
\Big[ -\frac{1}{2}\, \phi \, n^{-\frac{\Box}{2 m_n^2}} \, \phi +
\frac{1}{n + 1} \, \phi^{n+1} \Big]\,, \label{3.1}
\end{equation} whose explicit realization depends on particular
choice of coefficients $C_n$, string masses $m_n$ and coupling
constants $g_n$. To avoid a divergence  in $1/(n-1)$ when $n = 1$
one has to take that ${C_n\, m_n^D}/{g_n^2}$ is proportional to $n
-1$. Here we  consider some cases when coefficients $C_n$ are
proportional to $n-1$, while masses $m_n$ as well as coupling
constants $g_n$ do not depend on $n$, i.e. $ m_n =  m , \,\, g_n =
g$.  To differ this new field from a particular $p$-adic one, we
use notation $\phi$ instead of $\varphi$.

We have considered three cases for coefficients $C_n$ in
(\ref{3.1}): (i) $C_n = \frac{n-1}{n^{2+h}}$, where $h$ is a real
parameter; (ii) $C_n = \frac{n^2 -1}{n^2}$; and (iii) $C_n = \mu
(n)\, \frac{n-1}{n^2}$, where $\mu (n)$ is the M\"obius function.

Case (i) was considered in \cite{dragovich3}. Obtained Lagrangian
is
\begin{equation} L_{h} =  \frac{m^D}{g^2} \Big[ \,- \frac{1}{2}\,
 \phi \,  \zeta\Big({\frac{\Box}{2 \, m^2}  +
h }\Big) \, \phi    + {\cal{AC}} \sum_{n= 1}^{+\infty} \frac{n^{-
h}}{n + 1} \, \phi^{n+1} \Big]\,, \label{3.2} \end{equation} where
$\mathcal{AC}$ denotes analytic continuation.

Case (ii) was investigated in \cite{dragovich4} and the
corresponding Lagrangian is

 \begin{equation} L =  \frac{m^D}{g^2} \Big[ \, - \frac{1}{2}\,
 \phi \,  \Big\{ \zeta\Big({\frac{\Box}{2\, m^2}  -
 1}\Big)\, + \, \zeta\Big({\frac{\Box}{2\, m^2} }\Big) \Big\} \, \phi \,  + \,   \frac{\phi^2}{1 - \phi} \,
 \Big]\,. \label{3.3} \end{equation}

Case with the M\"obius function $\mu (n)$ is presented in
\cite{dragovich5}. Recall that the M\"obius function is defined
for all positive integers and has values $1, 0, -1$ depending on
factorization of $n$ into prime numbers $p$. Its explicit
definition as follows:
\begin{equation}
\mu (n)= \left \{ \begin{array}{lll} 0 , \quad &  n = p^2 m \,, \\
(-1)^k , \quad & n = p_1 p_2 \cdots p_k ,\,\,  p_i \neq p_j \,, \\
1 , \quad & n = 1, \,\,  (k=0)\, .
\end{array} \right.
\label{3.4}
\end{equation}
Since the inverse Riemann zeta function can be defined as
\begin{equation}
\frac{1}{\zeta (s)} = \sum_{n =1}^{+\infty}\, \frac{\mu (n)}{n^s},
\quad s=\sigma + i \tau , \quad \sigma > 1\,, \label{3.5}
\end{equation}
then the corresponding Lagrangian is

\begin{equation}
L =   \frac{m^D}{g^2} \Big[ - \frac{1}{2}\, \phi \, \frac{1}{
\zeta\Big({\frac{\Box}{2 m^2}}\Big)} \,\phi + \int_0^\phi {\cal
M}(\phi) \, d\phi\Big] , \label{3.6}
\end{equation}
where ${\cal M}(\phi) = \sum_{n= 1}^{+\infty} {\mu (n)} \,
\phi^{n} = \phi - \phi^2 - \phi^3 - \phi^5 + \phi^6 - \phi^7 +
\phi^{10} - \phi^{11} - \dots $.

\subsection{Multiplicative approach}

Let us now consider a new approach, which is not based on a
summation of $p$-adic Lagrangians, but the Riemann zeta function
will emerge through its  product form. Our starting point is again
$p$-adic Lagrangian  (\ref{2.1}) with equal masses, i.e. $m_p^2 =
m^2$ for every $p$. It is useful to rewrite (\ref{2.1}), first in
the form,
\begin{equation} {\cal L}_p = \frac{m^D}{g_p^2}\, \frac{p^2}{p^2-1} \Big\{
-\frac{1}{2}\, \varphi \, \Big[ p^{-\frac{\Box}{2 m^2}+1} +
p^{-\frac{\Box}{2 m^2}} \Big]\, \varphi + \, \varphi^{p+1}
\Big\}\, \label{3.2.1}
\end{equation}
and then, by addition  and substraction of $\varphi^2$, as
\begin{equation} {\cal L}_p = \frac{m^D}{g_p^2}\, \frac{p^2}{p^2-1} \Big\{
\frac{1}{2}\, \varphi \, \Big[ \Big(1 - p^{-\frac{\Box}{2 m^2}+1}
\Big) + \Big( 1 - p^{-\frac{\Box}{2 m^2}}\Big) \Big]\, \varphi -
\varphi^2 \Big( 1 - \varphi^{p-1} \Big) \Big\}\,. \label{3.2.2}
\end{equation}

Now we introduce a Lagrangian for the entire $p$-adic sector by
taking products
\begin{equation}
 \prod_p g_p^2  = C \,, \quad \prod_p \frac{1}{1 - p^{-2}}\,, \quad \prod_p (1 - p^{-\frac{\Box}{2 m^2}+1}) \,,
 \quad \prod_p (1 - p^{-\frac{\Box}{2 m^2}}) \,\, \mbox{and} \,\, \prod_p ( 1 - \varphi^{p-1})  \label{3.2.3}
\end{equation}
in (\ref{3.2.2}) at the corresponding places. Then this new
Lagrangian becomes
\begin{equation}
{\mathcal L} = \frac{m^D}{C}\, \zeta (2)\, \Big\{ \frac{1}{2} \,
\phi \Big[ \zeta^{-1} \Big( \frac{\Box}{2 m^2} - 1 \Big) +
\zeta^{-1} \Big( \frac{\Box}{2 m^2}  \Big)\Big] \, \phi - \phi^2
\prod_p \Big( 1 - \phi^{p-1} \Big) \Big\} \,,  \label{3.2.4}
\end{equation}
where $\zeta^{-1} (s) = 1/\zeta (s)$ and new scalar field is
denoted by $\phi$. For the coupling constant $g_p$ there are two
interesting possibilities: (1) $g_p^2 = \frac{p^2}{p^2 - 1}$, what
yields $\zeta (2) / C = 1$ in (\ref{3.2.4}), and (2) $g_p =
|r|_p$, where $r$ may be any non zero rational number and it gives
$|r|_\infty \prod_p |r|_p = 1$ (this possibility was considered in
\cite{dragovich6}). Both  these possibilities are consistent with
adelic product formula (\ref{1.6}). For simplicity, in the sequel
we shall take $C = \zeta (2)$. It is worth noting that having
Lagrangian (\ref{3.2.4}) one can easily reproduce its $p$-adic
ingredient (\ref{3.2.1}). Namely, one can use the opposite
procedure to the construction of (\ref{3.2.4}) from (\ref{3.2.1}).

Let us rewrite (\ref{3.2.4}) in the simple form
\begin{equation}
{\mathcal L}\, = \, \frac{1}{2} \, \phi\, \Big[ \zeta^{-1} \Big(
\frac{\Box}{2 } - 1 \Big) + \zeta^{-1} \Big( \frac{\Box}{2 }
\Big)\Big] \, \phi - \phi^2 \, \Phi (\phi) \,, \label{3.2.5}
\end{equation}
with $m = 1$ and $\Phi (\phi) = \mathcal{AC} \prod_p (1
-\phi^{p-1})$, where $\mathcal{AC}$ denotes analytic continuation
of infinite product $\prod_p (1 -\phi^{p-1})$, which is convergent
if $|\phi|_\infty < 1 $. One can easily see that $\Phi (0) = 1$
and $\Phi (1) = \Phi (-1) = 0$.

For (\ref{3.2.5}), the corresponding equation of motion  is
\begin{equation}
 \Big[ \zeta^{-1} \Big( \frac{\Box}{2 } - 1 \Big) + \zeta^{-1}
\Big( \frac{\Box}{2 } \Big)\Big] \, \phi  = 2 \phi \, \Phi (\phi)
+ \phi^2 \, \Phi' (\phi) \,, \label{3.2.6}
\end{equation}
and has $\phi = 0$ as a trivial solution. In the weak-field
approximation ($\phi (x) \ll 1$), equation (\ref{3.2.6}) becomes
\begin{equation}
 \Big[ \zeta^{-1} \Big( \frac{\Box}{2 } - 1 \Big) + \zeta^{-1}
\Big( \frac{\Box}{2 } \Big)\Big] \, \phi  = 2 \phi \,.
\label{3.2.7}
\end{equation}
Note that the above operator-valued zeta function can be regarded
as a pseudodifferential operator. Then (\ref{3.2.6}) and
(\ref{3.2.7}) are transformed to the integral form.

Mass spectrum of $M^2$ is determined by solutions of equation
\begin{equation}
  \zeta^{-1} \Big( \frac{M^2}{2 } - 1 \Big) + \zeta^{-1}
\Big( \frac{M^2}{2 } \Big)   = 2 \,. \label{3.2.8}
\end{equation}
There are infinitely many tachyon solutions, which are below
largest one $M^2\approx - 3.5$.

The potential follows from  $ - {\mathcal L}$ at $\Box = 0$, i.e.
\begin{equation}
V (\phi) =  [ 7 + \Phi (\phi)] \, \phi^2 \,, \label{3.2.9}
\end{equation}
since $\zeta (- 1) = - 1/12$ and $\zeta (0) = - 1/2$. This
potential has local minimum  $V (0) = 0$ and values $V (\pm 1) =
7$. To explore behavior of $V (\phi)$ for all $\phi \in
\mathbb{R}$ one has first to investigate properties of the
function $\Phi (\phi)$.

\section{Concluding remarks}

We have obtained a new Lagrangian (\ref{3.2.5}),  which
incorporates  all $p$-adic Lagrangians (\ref{2.1}) by a
multiplicative way. It contains information on the entire $p$-adic
sector. To compare it with $p$-adic sector of the adelic  string
one has to derive four-point scattering amplitude. It will be
considered elsewhere. There are also many classical properties
which should be explored, e.g. possible solutions of the equation
of motion (\ref{3.2.6}).

It is worth noting that a Lagrangian similar to (\ref{3.2.5}) can
be obtained by an additive approach. Namely, let us start from
(\ref{3.2.1}) and take coupling constant $g_p^2 = \frac{p^2}{p^2 -
1}$ and mass $m = 1$. Then let us extend prime $p$ to natural
number $n$ and take infinite sum
\begin{equation}
L = - \sum_{n=1}^{+\infty} \mu (n) \, \mathcal{L}_n =
\frac{1}{2}\, \phi \, \Big[ \sum_{n=1}^{+\infty} \mu (n) \,
p^{-\frac{\Box}{2 } + 1}  + \sum_{n=1}^{+\infty} \mu (n) \,
p^{-\frac{\Box}{2 }} \Big] \, \phi  - \sum_{n=1}^{+\infty} \mu (n)
\, \phi^{n+1}\,, \label{4.1}
\end{equation}
where $\mu (n)$ is the above M\"obius function defined by
(\ref{3.4}). Introducing zeta function in (\ref{4.1}), one can
rewrite it in the form
\begin{equation} L  = \frac{1}{2}\, \phi\,
\Big[ \zeta^{-1} (\frac{\Box}{2 } - 1)  +  \zeta^{-1}
(\frac{\Box}{2 }) \Big] \, \phi -  \phi^{2}\, F (\phi) \,,
\label{4.2}
\end{equation}
where $F (\phi) = \sum_{n=1}^{+\infty} \mu (n) \, \phi^{n-1} $.
The difference  between (\ref{3.2.5}) and (\ref{4.2}) is only in
functions $\Phi (\phi)$ and $F (\phi)$. Since $\Phi (\phi) = (1 -
\phi) (1 - \phi^2) (1-\phi^4) ... = 1 - \phi -\phi^2 + \phi^3 -
\phi^4 + ...$ and $F (\phi) = 1 - \phi - \phi^2 - \phi^4 +...$, it
follows that these functions have the same behavior for $|\phi|
\ll 1$. Hence,   Lagrangians (\ref{3.2.5}) and (\ref{4.2}) have
the same mass spectrum and in weak-field approximation describe
the same scalar field theory.

Note that an interesting approach to the foundation of a field
theory and cosmology based on the Riemann zeta function was
proposed in \cite{volovich3}.

\bigskip
\bigskip

{\bf Acknowledgements}
\bigskip

  This paper was supported in part by the Ministry of Science and Technological Development, Serbia\,
  (Contract No. 144032D). The author would like to thank organizers of the 4-th
RTN ``Forces-Universe'' EU Network Workshop in Varna, 11-17
September 2008, for invitation to present preliminary results of
this work as well as for kind hospitality. The author also thanks
I. Ya. Aref'eva and I. V. Volovich for useful discussions.

%The style of the following references should be used in all documents.

\end{document}